# Modeling transport through single-molecule junctions

Kamil Walczak [1,2] and Sergey Edward Lyshevski [3]

[1] Institute of Physics, A. Mickiewicz University, Umultowska 85, 61-614 Poznań, Poland

[2] Instituto de Ciencia de Materiales de Madrid (CSIC), Cantoblanco, 28049 Madrid, Spain

[3] Department of Electrical Engineering, Rochester Institute of Technology,
14623-5603 New York, USA

**Abtract.** Non-equilibrium Green's functions (NEGF) formalism combined with extended Hückel (EHT) and charging model are used to study electrical conduction through single-molecule junctions. The analyzed molecular complex is composed of the asymmetric 1,4-Bis((2'-*para*-mercaptophenyl)-ethinyl)-2-acetyl-amino-5-nitro-benzene molecule symmetrically coupled to two gold electrodes [Reichert *et al.*, Phys. Rev. Lett. Vol. 88, (2002), pp. 176804]. Owing to this model, the accurate values of the current flowing through such junctions can be obtained by utilizing basic fundamentals and coherently deriving model parameters. Furthermore, the influence of the charging effect on the transport characteristics is emphasized. In particular, charging-induced reduction of conductance gap, charging-induced rectification effect and charging-generated negative value of the second derivative of the current with respect to voltage are observed and examined for the molecular complex.

Key words: transport, rectification, charging, molecular electronics, molecular junction
PACS numbers: 73.23.-b, 85.65.+h

## 1. Introduction

In recent years, a few experiments on molecular-scale juncions were performed and their transport characteristics were successively measured [1-5]. Such junctions are usually composed of individual molecules (or molecular layers) connected to two (or more) metallic electrodes, operating under the influence of a bias voltage. Since that time molecular electronics has become very popular among scientists [6-11], offering an opportunity to connect particular devices into the properly working integrated circuit (nanoIC). That is why the understanding of conduction at the molecular scale and modeling of electrical transport through molecule-based junctions are of great interest nowadays. In particular, a great challenge is to reproduce experimental data with the help of a theoretical description of the transport process [12]. Theoretically, current-voltage characteristics of molecular junctions were calculated using as well semi-empirical (parametric) [1,3,13-15] as *ab initio* (first-principles) [12,16-21] methods. Furthermore, from the technological viewpoint it is very important to control the molecule-to-electrodes contacts and to guarantee chemical stability (such instability is associated with voltages higher than 1 Volt [4]). Still open question remains the appearance of time-dependent phenomena in molecular-scale devices, such as: single-electron pumps and turnstiles, as-field response and transients in resonant-tunneling, and interaction with ultrashort laser pulses [22].

In this work we focus our attention on qualitative description of transport through molecule-based junctions. Calculations of transport characteristics were performed using Extended Hückel Theory (EHT) and simplified charging model in the computational scheme based on Non-Equilibrium Green Function (NEGF) formalism [23]. EHT method is widely used in order to describe the electronic structure of molecules and uses all the valence orbitals of the atoms as the non-orthogonal basis functions, neglecting all the interactions between electrons (which is a strong defect of this type of modeling). Energy levels obtained within EHT method are lowered by a few eV in comparison with the correct values relative to a vacuum, but energy differences are usually comparable with *ab initio* calculations. Anyway, this approach provides reasonably accurate quantitative results that ensure insight into the basic physics of analyzed problem.

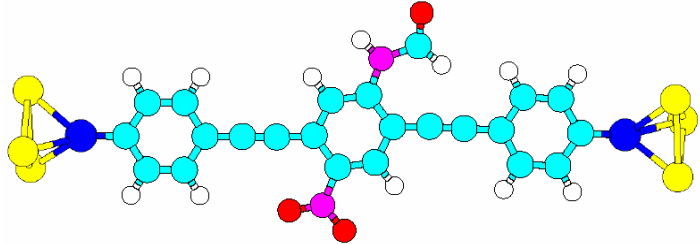

Figure 1  Schematic representation of the analyzed molecule connected to the electrodes
(C: cyan, N: violet, O: red, S: blue, H: white, Au: yellow).

**2. Molecular complexes and computational scheme**

As an example we have studied the asymmetric 1,4-Bis((2'-*para*-mercaptophenyl)-ethinyl)-2-acetyl-amino-5-nitro-benzene molecule symmetrically coupled to two gold electrodes [4], as shown in Figure 1. Gold electrodes Au(111) are comonly used to form good contacts with thiol end groups (-SH). In the first step of the computational scheme, the geometry of the isolated molecule has been optimized with the help of Parametric Method 3 (PM3) as implemented in the HyperChem package [24]. Calculations at this level confirmed that the considered molecule is planar. Then the end hydrogen atoms were removed, as they are lost by thiol groups in the chemisorption process, when molecule is attached to the gold pads [25]. Transport characteristics were obtained for the molecule connected to the gold electrodes on both sides. As a part of the so-called extended molecule three gold atoms in the 111 surface were added to both ends of the molecule (sulfur atoms), where the distance between sulfur terminal atoms and gold surfaces is 1.905 Å (which corresponds to a S-Au bond length of 2.53 Å with the angle 41 degrees). The Au-Au bond length is 2.885 Å. The rest of the gold atoms are included into the computational scheme through self-energy terms, where the surface Green functions are computed exactly by making use of the periodicity of the semi-infinite electrodes [23].

In the NEGF formalism, the current flowing through the device is calculated with the help of the standard Landauer-type formula:

$$I(V) = \frac{2e}{h} \int_{-\infty}^{+\infty} Tr(\Gamma_1 G \Gamma_2 G^+)[f(\omega-\mu_1) - f(\omega-\mu_2)]d\omega, \qquad (1)$$

where: broadening matrices $\Gamma_{1,2} = i(\Sigma_{1,2} - \Sigma_{1,2}^+)$ are defined as the anti-Hermitian parts of self-energies $\Sigma_{1,2}$. The molecular Green function is given by: $G = (\omega S - H + U_{SCF} - \Sigma_1 - \Sigma_2)$, where:

$S$ is the overlap matrix for non-orthogonal basis set of states, while $H$ is the Hamiltonian of the molecule. In Eq.1 $f(\omega-\mu_{1,2})$ denote the Fermi distributions with electrochemical potentials defined as: $\mu_{1,2}=\varepsilon_F \pm eV/2$. Semi-empirical methods like EHT use simplified procedures in order to impose self-consistent potential $U_{SCF}$. Non-equilibrium transport characteristics were calculated by employing a simple Self-Consistent Field (SCF) method modeling SCF potential by the term of $U_{SCF}=U(N-N_{eq})$, where $N_{eq}$ is equilibrium number of electrons in the molecule $N_{eq}=144$, while the number of electrons on the molecule is calculated using the following integral [23]:

$$N = \frac{1}{2\pi}\int_{-\infty}^{+\infty} d\omega \left[ f(\omega-\mu_1)Tr(G\Gamma_1 G^+) + f(\omega-\mu_2)Tr(G\Gamma_2 G^+) \right]. \quad (2)$$

This means that the potential profile inside the molecule is assumed to be flat depending on charge energy $U$ only.

Nonlinear transport characteristics of molecular junctions strongly depend not only on the potential profile across the molecule under the influence of a bias voltage, but also on the location of Fermi energy with respect to energy levels of the molecule. Fermi energy is considered as a fitting parameter within reasonable limits (here we assumed that $\varepsilon_F=-11.000$ eV, which is in good agreement with the values known from the literature). However, Fermi energy location can be estimated by a formal requirement that the number of states below the Fermi energy must be equal to the number of electrons in the extended molecule [1]. The differential conductance is calculated from the current (Eq.1) as its derivative with respect to voltage. Finally, density of states is calculated as: $DOS=iTr[(G-G^+)S]/2\pi$, while the transmission function is given through the relation: $T=Tr(\Gamma_1 G \Gamma_2 G^+)$. Furthermore, all the calculations were performed at the room temperature (0.025 eV), while the extended molecule is taken into account only for calculations at equilibrium conditions (i.e. DOS and transmission spectra). The second statement is due to the fact that constant potential across the central part of the device is better justified for the non-extended molecule, since the cluster of metal atoms at each end of the device tends to have the same potential as the bulk contact.

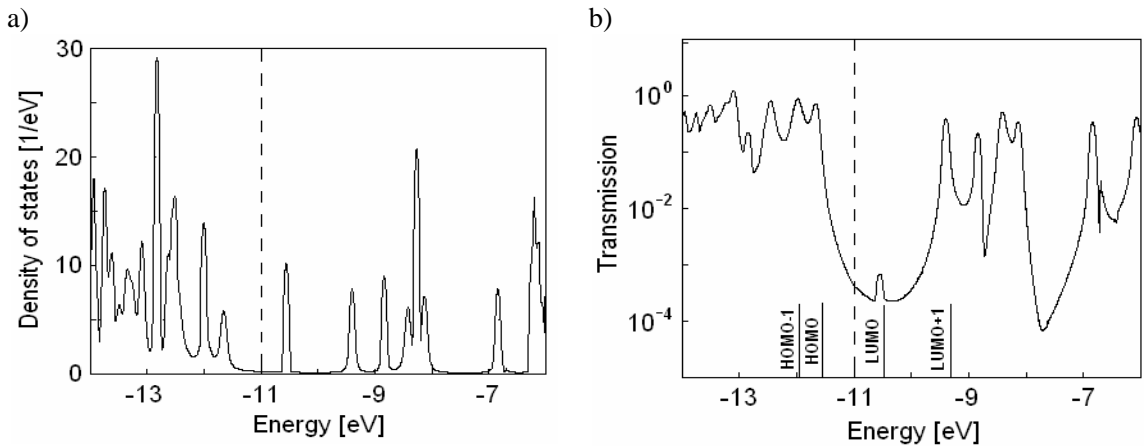

Figure 2  Total density of states (a) and transmission (b) as a function of electron energy for the analyzed molecular junction. The dashed line denotes the Fermi energy.

## 3. Results and discussion

Figure 2a presents the total density of states (DOS) as a function of electron energy (the dashed line denotes the Fermi energy), where the coupling with the electrodes results in the broadening and shifting of the discrete energy levels of the molecule. The transmision probability as a function of electron energy is reported in Figure 2b and more or less resembles the total DOS. However, it is interesting why in the DOS we observe a distinct peak for the LUMO level, while transmission through this level is very small (three orders of magnitude smaller than the unit transmission). Such behavior can be justified as a consequence of quantum interference of the molecular orbitals, where the variation of interference conditions is due to the presence of chemical substituent groups [26]. Although the strong chemical bond is realized between the sulfur atoms and gold pads, both DOS and transmission functions are sharply peaked around the molecular energy levels. This reflects the localized character of the chemical bond, which means that the sulfur atoms partially insulate the delocalized pi-electron states which are mainly involved into the conduction process (it is a weak-coupling case in which conducting channels are electrically isolated from metallic electrodes via potential barriers).

So a molecular island can be treated conceptually as a molecular quantum dot and therefore our simplified charging model is somehow adequate. However, in reality, the electrostatic potential profile could be more complicated than the assumed flat potential along the molecule [27]. Anyway, the problem we are facing now is to evaluate the charging energy of the capacitor formed by molecule coupled via tunnel barriers with capacitances $C_1$ and $C_2$ to the source and drain electrodes. The stored electrostatic energy of this capacitor is expressed as: $U = e^2/(2C)$, where $e$ is an electron charge, while $C = C_1 + C_2$ is total capacitance of the system. Since the capacitance in organic molecular junctions can reach the value of $C \sim 10^{-19}$ F [2], charging energy corresponds to $U \sim 1$ eV. Moreover, it is rather clear that the charging energy $U$ is influenced mainly by the extent of the electronic wavefunction. Generally, the more localized the wavefunction, the smaller value of the $U$-parameter.

The calculated accurate current-voltage (I-V: solid lines) and conductance-voltage (G-V: dshed lines) characteristics are presented in Figure 3 for three different charge energies: $U = 0, 1$ and 2 eV, respectively. Since the height of the LUMO peak in the transmission function is negligibly small, one can expect no evidence of that peak in the transport dependences. This is a case only in the absence of charging, see Figure 3a. However, when the charge energy is not equal to zero, the LUMO peak is evident as a jump in the G-V characteristic, as presented in Figures 3b and 3c. Furthermore, our treatment of charging results in the reduction of the conductance gap (CG) with an increase of the $U$-parameter. The explanation of such behavior is the following: as charging is increased, molecular electronic structure is shifted in relation to Fermi energy and therefore LUMO level reaches the resonant condition at lower voltages. Here the conductance gap is determined mainly by the energy difference between the Fermi energy and the LUMO level of the molecular system (CG is also slightly affected by the strength of the molecule-to-electrodes coupling and temperature). By the way, it should also be mentioned that some state-of-art first-principles calculations overestimate the CG quantity in comparison with experimental data [12]. Here we indicate that charging has a great influence on determination of conductance gap.

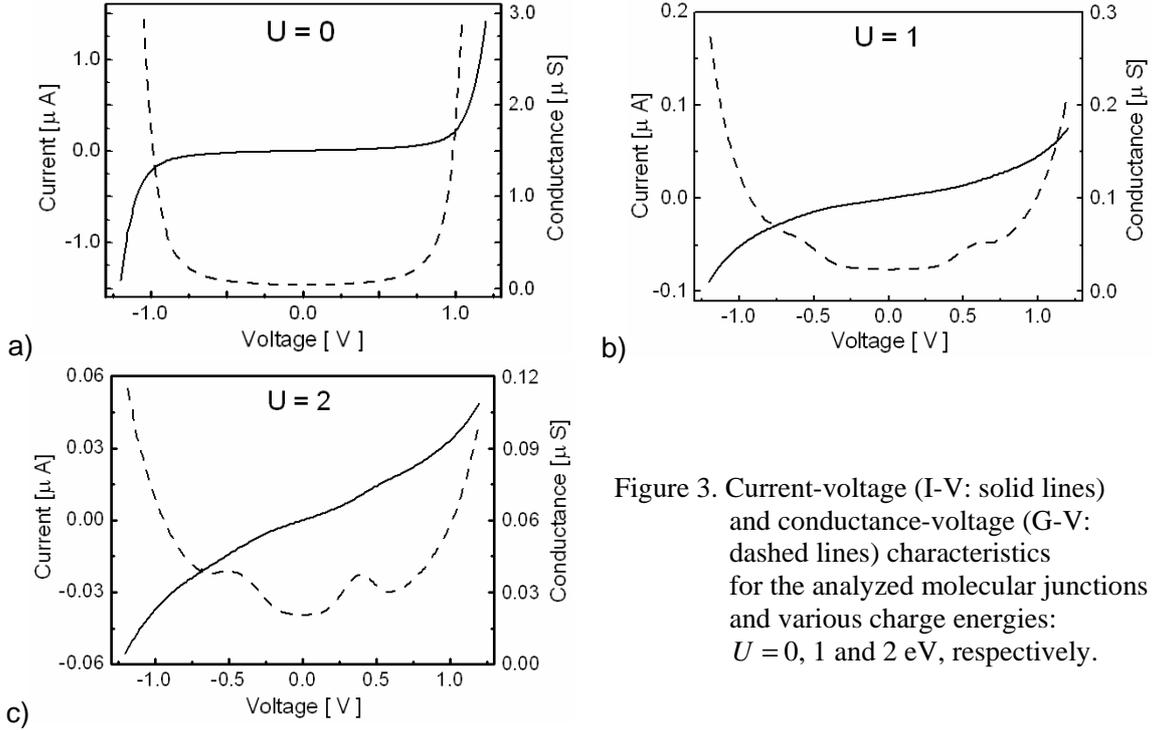

Figure 3. Current-voltage (I-V: solid lines) and conductance-voltage (G-V: dashed lines) characteristics for the analyzed molecular junctions and various charge energies: $U = 0$, 1 and 2 eV, respectively.

Our results also indicate the effect of $U$-induced rectification, since diode-like behavior of the asymmetric molecular complex is not observed in the case when charging is completely neglected (see Figure 3a). Within our model, the mentioned phenomenon is a combined effect of the asymmetric structure of molecule itself and the charging profile. Current rectification is due to the presence of chemical substituent groups and their influence on charge redistribution within the molecular system. Since the location of substituents is spatially asymmetric with respect to the direction of the current flow, the response upon the applied bias is also asymmetric in the case of non-zero charging. The general tendency is the following: the bigger the value of the charging $U$-parameter, the stronger the asymmetry of transport characteristics associated with the change of bias polarity. Moreover, for the extremely large charging parameter $U = 2$ eV the second derivative of the current with respect to voltage is negative, i.e. the differential conductance decreases with increasing bias voltage (as viewed from Figure 3c for $V > 0.4$ Volt). It should also be noted by the reader that the scales of current and conductance are slightly different for different $U$-parameters for the analyzed range of bias voltages (compare Figures 3a, 3b and 3c). The reason is associated with the fact that the effect of charging is to smooth I-V characteristics and to reduce the height of conductance peaks as well as to broaden the peaks in conductance spectra [28].

### 4. Concluding remarks

In summary, we have introduced calculations of transport characteristics through concrete molecular junction. Such calculations were based on NEGF formalism in combination with EHT description of the molecular system and simplified charging model to include potential changes inside the molecule. Although this approach provides reasonably accurate quantitative results (comparable with experimental data), it should be noted that the method is based on a few crucial approximations. Transport considered in this work is assumed to be elastic and coherent, which means that our computational scheme neglects all inelastic and incoherent scattering processes inside the molecule. As was mentioned before, at the EHT level of molecular description all the effects associated with electron correlations are ignored.

Moreover, structural changes of the molecule under the influence of a bias voltage are beyond the scope of this work. Although it is possible to calculate the optimal geometry at equilibrium (at zero bias), the simple variational principle is no longer valid in the presence of current flow. It means that we can not minimize energy by varying the geometry of the molecular bridge and therefore the assumed geometries are used instead. But anyway, some first-principles calculations based on the Hellmann-Feynman type of theorem show that current-induced forces on small molecules have negligible influence on transport in the range of voltages considered in this work [29].

It should also be mentioned that transport characteristics for the analyzed device were calculated within a more advanced density functional theory (DFT) by Heurich *et al.*, where the effects of electron-electron interactions are taken into account in the self-consistent way [21]. However, DFT calculations are adequate for basic states, however in non-equilibrium conditions excited states also play an important role. Moreover, for the case of an asymmetric molecule the experimental data show asymmetric transport characteristics and such asymmetry is almost invisible in the cited work [21]. In the present paper we show that charging effects are responsible for the asymmetry of transport characteristics.

**Acknowledgements**

K.W. is very grateful to G. Platero, A. Lehmann-Szweykowska and T. Kostyrko for many enlightening discussions. This work was supported by the European Commission contract No. HPRN-CT-2002-00282.